\documentstyle[prl,twocolumn,aps]{revtex}

  \input epsf
  \begin{document}
  \draft
  \title{Coherent acceleration and Landau-Zener tunneling of Bose-Einstein
  condensates in 1-D optical lattices}
  \author{O. Morsch, J.H. M\"uller, M. Cristiani,
  and E. Arimondo}
  \address{INFM, Dipartimento di Fisica, Universit\`{a} di Pisa, Via
  Buonarroti 2, I-56127 Pisa, Italy}
  \date{\today}
  \maketitle
  \begin{abstract}
  We have loaded Bose-Einstein condensates into one-dimensional, off-resonant
  optical lattices and
  accelerated them by chirping the frequency difference between the two
  lattice beams. For small
  values of the lattice well-depth, Bloch oscillations were observed.
  Landau-Zener tunneling out of
  the lowest lattice band, leading to a breakdown of the oscillations, was
  also studied. In order to
  allow in-trap measurements of these phenomena, we dynamically compensated
  for the intrinsic
  micromotion of the atoms in our time-orbiting potential trap.
  \end{abstract}

  \narrowtext

  The properties of ultra-cold atoms in periodic light-shift potentials in
  one, two and
  three dimensions have been investigated extensively in the past ten
  years~\cite{lattreview}. In
  near-resonant and, more recently, far-detuned optical lattices, a variety
  of phenomena have been
  studied, such as the magnetic properties of atoms in optical lattices,
  revivals of wave-packet
  oscillations, and Bloch oscillations in accelerated
  lattices~\cite{raizen97}.
  While in most of the original optical lattice experiments the atomic clouds
  had temperatures in the
  the micro-Kelvin range, corresponding to a few recoil energies of the
  atoms, atomic samples with
  sub-recoil energies are now routinely produced in Bose-Einstein
  condensation experiments. Since the
  first experimental realizations in 1995, many aspects of Bose-Einstein
  condensed atomic clouds
  (BECs) have been studied~\cite{inguscio}, ranging from collective
  excitations to superfluid
  properties and quantized vortices. So far, the majority of these
  experiments have been carried out
  essentially in harmonic-oscillator potentials provided by magnetic traps or
  optical dipole traps.
  The properties of BECs in periodic potentials constitute a vast new field of
  research, for

instance~\cite{sorensen98,jaksch98,choi99,javanainen99,brunello00,zobay00,chiofalo00,trombettoni01,poetting01}.
  Several experiments in the pulsed standing wave
  regime~\cite{kozuma99,stenger99} as well as studies of the
  tunneling of BECs out of the potential wells of a shallow optical lattice
  in the presence of
  gravity~\cite{anderson98}, and recent search for the superfluid
  dynamics\cite{burger01} have taken the first steps in that direction. In
  this paper, we present
  the results of experiments on BECs of $^{87}\mathrm{Rb}$ atoms in
  accelerated optical lattices. In
  particular, we demonstrate coherent acceleration of BECs adiabatically
  loaded into optical
  lattices. For small values of the lattice depth we observed Bloch
  oscillations, which exhibited
  Landau-Zener breakdown when the lattice depth was further reduced and/or
  the acceleration
  increased. We loaded the condensate into optical lattices with different
  spatial
  periods, generating the periodic optical lattice either from two
  counterpropagating laser beams
   or two laser beam enclosing an angle $\theta$ different from $180$ degrees.

  The properties of a Bose-Einstein condensate located in a periodic optical
  lattice with depth $U_0$ are described through the Gross-Pitaevskii
  equation valid for the
  single-particle wavefunction.
  Because the nonlinear term in the Gross-Pitaevskii equation reproduces the
  spatial periodicity of the
  wavefunction, the condensate ground state is periodic. In agreement
  with the Bloch approach, the condensate excitation spectrum exhibits a
  band structure and in presence of an
  acceleration of the optical lattice, Bloch oscillations of the condensate
  should occur
  \cite{sorensen98,choi99}. We present experimental results for Bloch
  oscillations preserving the condensate wavefunction. The role of the
  nonlinear interaction
  term of the Gross-Pitaevskii equation may be
  described through an effective potential in a noninteracting gas model
  \cite{choi99,chiofalo00}.
  In the perturbative regime of ref.~\cite{choi99} the effective potential
  is $U_{\rm eff}=U_0/(1+4C)$ with $C=g/E_{\rm B}$ the interaction ratio
  between the nonlinear interaction
  term $g=4\pi n_0 \hbar^2 a/M$ and the lattice Bloch energy
  $E_{\rm B}=\hbar^2(2\pi)^2/Md^2$. The parameter $C$ contains the
  condensate density $n_0$, the s-wave scattering
  length $a$, the atomic mass $M$, the lattice constant $d=\pi
  /\sin(\theta/2)k$,
  with $k$ the laser wavenumber, and $\theta$ the angle between the two
  laser beams creating the
  1-D optical lattice. From this it follows that a small angle $\theta$
should result in a large
  interaction term $C$. In the following, the parameters $d$,
  $E_{\rm B}$ and $C$ always refer to the respective lattice
  geometries with angle $\theta$.

  Our apparatus used to achieve Bose-Einstein condensation of
  $^{87}\mathrm{Rb}$ is described in detail in~\cite{jphysbpaper}.
  Essentially, $~5\times
  10^7$ atoms captured in a magneto-optical trap (MOT) were transferred into
  a triaxial
  time-orbiting potential trap (TOP)~\cite{prlpaper}. Subsequently, the atoms
  were evaporatively
  cooled down to the transition temperature for Bose-Einstein condensation,
  and after further cooling
  we obtained condensates of $\approx 10^4$ atoms without a discernible thermal
  component in a magnetic trap with frequencies around
$15-30\,\mathrm{Hz}$. In one
  set of experiments, the magnetic trap was then switched off and a
  horizontal 1-D optical lattice was switched on, while in the other case the
  interaction between the
  condensate and the lattice took place inside the magnetic trap, which was
  subsequently switched off
  to allow time-of-flight imaging. The lattice beams were created by a
  $50\,\mathrm{mW}$ diode
  slave-laser injected by a grating-stabilized master-laser blue-detuned by
  $\Delta\approx
  28-35\,\mathrm{GHz}$ from the $^{87}\mathrm{Rb}$ resonance line. After
  passage through an optical
  fibre, the laser light was split and passed through two acousto-optic
  modulators (AOMs) that were
  separately controlled by two phase-locked RF function generators operating
  at frequencies around
  $80\,\mathrm{MHz}$, with a frequency difference $\delta$. The first-order
output beams of the
  AOMs generated the
  optical lattice. An acceleration of the lattice was effected by applying
  a linear ramp to $\delta$. For the values of the detuning and
  laser
  intensity used in our
  experiment, the spontaneous photon scattering rate ($\approx
  10\,\mathrm{s^{-1}}$) was negligible
  during the interaction times of a few milliseconds. In our experimental
  setup, we
  realized a counter-propagating lattice geometry with
  $\theta=180\,\mathrm{deg}$
   and an angle geometry with $\theta=29\,\mathrm{deg}$, leading to
   lattice constants $d$ of  $0.39$ and 1.56 $\mathrm{\mu m}$, respectively.
   The typical condensate density of $n_0\approx 5\times
  10^{13}\,\mathrm{cm^{-3}}$ for our trap parameters in the Thomas-Fermi
limit leads to $C=0.01$
  for the counterpropagating configuration and $C=0.17$ for
$\theta=29\,\mathrm{deg}$~\cite{footnote}. We therefore expect mean-field
effects to
  be negligible
  in the counter-propagating lattice geometry, whereas in the geometry
  with a larger lattice
  constant the effective potential acting on the condensate should be
  significantly
  reduced. The two lattice beams with $~3\,\mathrm{mW}$ each were expanded to a
  waist of
  $1.8\,\mathrm{mm}$, giving a
  theoretical maximum lattice depth (for $\Delta\approx 28\,\mathrm{GHz}$) of
  $U_0\approx 4\,E_{\rm B}$ for the counter-propagating
  lattice geometry.

  \begin{figure}
  \centering\begin{center}\mbox{\epsfxsize 2.8 in \epsfbox{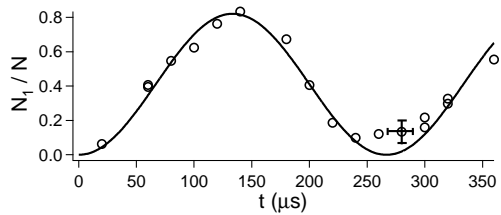}}
  \end{center}
  \caption{Rabi oscillations between two momentum states of a Bose-Einstein
  condensate. Shown here is
  the fraction $N_1/N$ of atoms in the $|p=2\hbar k\rangle$ momentum state
  as a function of time. From the measured Rabi frequency $\Omega_R/2\pi\approx
  3.6\,\mathrm{kHz}$
  we determined the lattice depth $U_0=2\hbar \Omega_R\approx 0.26\,E_{\rm
B}$.}
  \label{fig1}
  \end{figure}

  In a preliminary experiment aimed at determining the depth
  of the periodic optical potential, we flashed on the counterpropagating
  lattice with
  $\delta=\hbar k^{2}/(2\pi M)=15.08\,\mathrm{kHz}$ for
  $10-400\,\mathrm{\mu s}$. This detuning
  corresponds to the first Bragg resonance, causing the condensate to undergo
  Rabi oscillations
  between the momentum states $|p=0\rangle$ and $|p=2\hbar k\rangle$ (see
  Fig.~\ref{fig1}). From the
  measured Rabi frequency we could then determine the lattice
  depth~\cite{bendahan96}. The results of
  those measurements fell short by a factor of about 2 of the theoretical
  prediction, which was mainly
  due to the $10-15\%$ uncertainty
  in our intensity measurements and imperfections in the alignment and
  polarization of the lattice beams. As shown above, mean-field
  effects are expected to be negligible in this geometry.

  \begin{figure}
  \centering\begin{center}\mbox{\epsfxsize 2.8 in \epsfbox{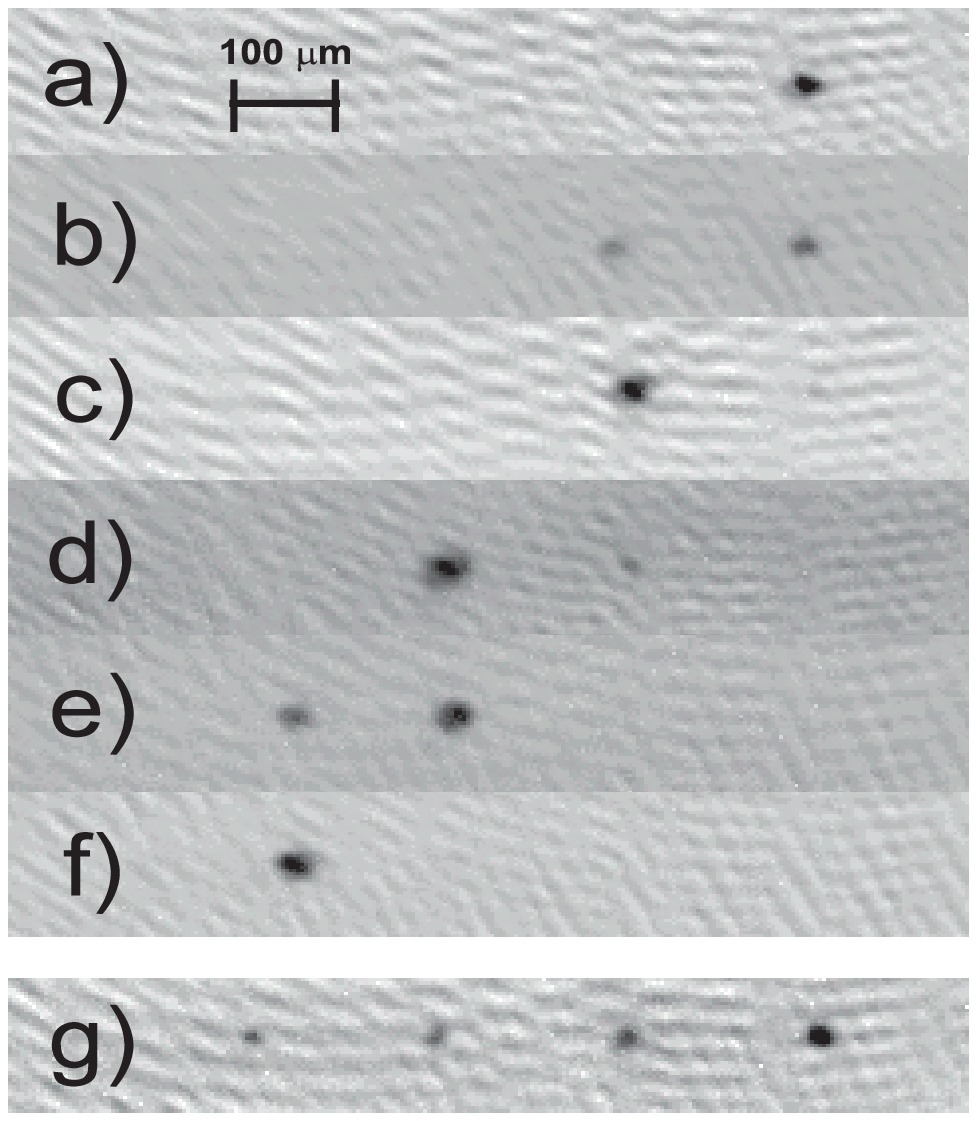}}
  \end{center}
  \caption{Coherent acceleration of a Bose-Einstein condensate. In (a)-(f)
  $U_0=0.29\,\mathrm{E_{\rm B}}$
  $a=9.81\,\mathrm{m\,s^{-2}}$, the condensate accelerated for
  $0.1,0.6,1.1,2.1,3.0$
  and $3.9\,\mathrm{ms}$, respectively. In (g) the result of
  $2.5\,\mathrm{ms}$ acceleration with the
  same lattice depth as above, but with $a=25\,\mathrm{m\,s^{-2}}$. In this
  case, a fraction of the
  condensate undergoes Landau-Zener tunneling out of the lowest band each
  time a Bragg-resonance is
  crossed. The separations between the different spots vary because detection
  occured after different
  time delays.}
  \label{fig2}
  \end{figure}

\begin{figure}
  \centering\begin{center}\mbox{\epsfxsize 2.8 in \epsfbox{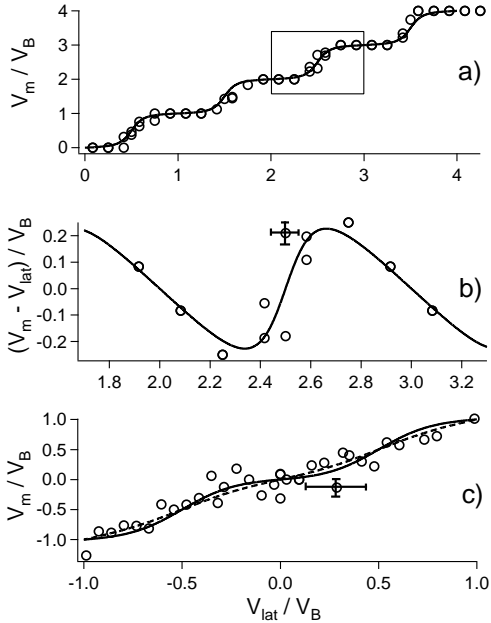}}
  \end{center}
  \caption{Bloch oscillations of the condensate mean
  velocity $v_{m}$ in an optical
  lattice. (a) Acceleration
  in the counterpropagating lattice with $d=0.39\,\mathrm{\mu m}$, $U_0\approx
  0.29\,E_{\rm B}$ and
  $a=9.81\,\mathrm{m\,s^{-2}}$. Solid line: theory. (b) Bloch oscillations
  in the rest frame of the
  lattice, along with the theoretical prediction (solid line) derived from
  the shape of the lowest
  Bloch band. (c) Acceleration in a lattice with $d=1.56\,\mathrm{\mu
  m}$,
  $U_0\approx
  1.38\,E_{\rm B}$ and $a=0.94\,\mathrm{m\,s^{-2}}$. In this case, the
Bloch oscillations are much less pronounced.
  Dashed and solid lines:
  theory for $U_0=1.38\,E_{\rm B}$ and $U_{eff}\approx 0.88\,E_{\rm B}$.}
  \label{fig3}
  \end{figure}

  In order to accelerate the condensate,
   we adiabatically loaded it into the
  lattice by switching one of the lattice beams on suddenly and ramping the
  intensity of the other
  beam from $0$ to its final value in $~200\,\mathrm{\mu s}
  $~\cite{footnote1}. Thereafter, the
  linear increase of the detuning $\delta$ provided a constant acceleration
  $a=\frac{\lambda}{2\sin(\theta/2)}\frac{d\delta}{dt}$ of the optical
lattice, leading to a final lattice velocity
  $v_{lat}=\frac{\lambda}{2\sin(\theta/2)}\delta$, where $\delta$ is the
final detuning between the beams. After a few
  milliseconds of acceleration, the lattice beams were switched off and the
  condensate was imaged
  after another $10-15\,\mathrm{ms}$ of free fall. As the lattice can
  only transfer momentum to the
  condensate in units of the Bloch momentum $p_{\rm B}=\hbar (2\pi/d)$, the
  acceleration of the condensate showed up as diffraction
  peaks corresponding to higher momentum classes as time increased
  (Fig.~\ref{fig2}). Since for our
   magnetic trap parameters the initial momentum spread of the condensate
(which is transferred into a spread of the lattice quasimomentum
   during an adiabatic switch-on) was
  much less than a
  recoil momentum of the optical lattice, the different
  momentum classes
  $|p=\pm np_{\rm B}\rangle$ (where $n=0,1,2,...$) occupied by the
  condensate wavefunction could be resolved directly after the
  time-of-flight. The average velocity of
  the condensate was derived from the occupation numbers of the different
momentum
  states. Figure~\ref{fig3}
  shows the results of the acceleration of a
  condensate  in the counterpropagating lattice with $U_{\rm
eff}=0.29\,E_{\rm B}$ and
  $a=9.81\,\mathrm{m\,s^{-2}}$. In the rest-frame of the
  lattice (Fig.~\ref{fig3} (b)), one clearly sees Bloch oscillations of the
  condensate velocity
  corresponding to a Bloch-period
  $\tau_{B}=\frac{h}{M_{Rb}ad}=1.2\,\mathrm{ms}$. The shape of
  these oscillations agrees well with the theoretical curve calculated from
  the lowest energy band of
  the lattice. As described in~\cite{bendahan96}, the acceleration
  process within a periodic potential can also be
  viewed as a succession of adiabatic rapid passages between momentum states
  $|p=2n\hbar k \rangle$
  and $|p=2(n+1)\hbar k \rangle$. We observed a momentum transfer
  of up to $6p_B$ without a detectable reduction of the
  phase-space density of the condensate~\cite{footnote2}.

\begin{figure}
  \centering\begin{center}\mbox{\epsfxsize 2.8 in \epsfbox{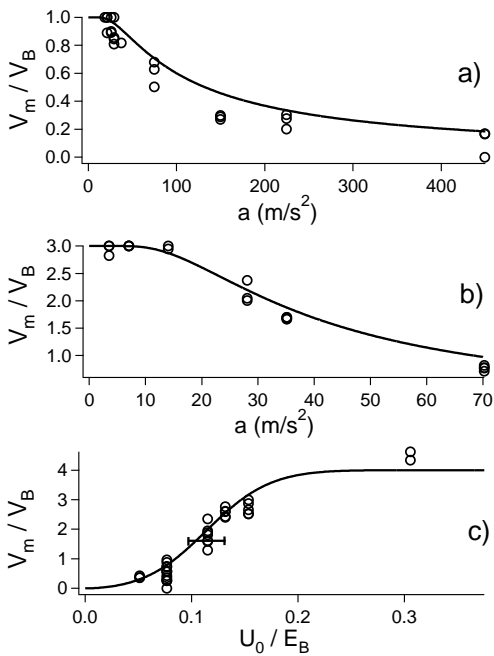}}
  \end{center}
  \caption{Landau-Zener tunneling of a condensate in an optical lattice. (a)
  and (b): Mean velocity
  of the condensate after acceleration of the lattice to $ v_{\rm B}$ and
  $3 v_{\rm B}$, respectively, as
  a function of acceleration. (c) Mean velocity of the condensate after
  acceleration of the lattice
  to $4.4v_{\rm B}$ as a function of lattice depth. In (a) and (b), the lattice
  depth was fixed at
  $U_0=0.25\,E_{\rm B}$, and in (c) the acceleration
  $a=9.81\,\mathrm{m\,s^{-2}}$. Solid lines: theory
  (see text). In (c), agreement with theory is expected to be somewhat less good because
  the final velocity of the lattice is not an integer multiple of $v_{\rm B}$ (see text).}
  \label{fig4}
  \end{figure}

  When we increased the acceleration of the lattice
  or decreased the lattice depth, not all of the condensate was coherently
  accelerated up to the
  final velocity of the lattice. This can be interpreted in
  terms of Landau-Zener
  tunneling of
  the condensate out of the lowest band when the edge of the Brillouin zone
  is reached. Each time the
  condensate is accelerated across this edge, a fraction $r=\exp(-a_c/a)$
  with $a_c=\frac{\pi
  U_0^2}{8\hbar p_{B}}$ undergoes tunneling into the first excited band (and,
  therefore, effectively
  to the continuum, as the gaps between higher bands are negligible for the
  shallow potentials used
  here). In Fig.~\ref{fig4} (a), the average velocity of the condensate
  after acceleration of
  the counter-propagating lattice to the Bloch velocity $v_{\rm B}=p_{\rm
  B}/M$ is shown as a function of acceleration and
  lattice depth along with a theoretical prediction using the Landau-Zener
  tunneling probability,
  which gives a mean velocity $v_m=(1-r)v_{\rm B}$ of the condensate for a
  final velocity $v_{\rm B}$ of
  the lattice. When the lattice is accelerated up to a velocity
  $nv_{\rm B}$, a straightforward
  generalization of this equation yields $v_m=v_{\rm B}(1/r-1)[1-(1-r)^n]$
  for the mean velocity of
  the condensate, which is in good agreement with our experimental findings
  (Fig.~\ref{fig4}
  (b))~\cite{footnote3}. In this case, a fraction $r$ of the condensate
undergoes Landau-Zener
  tunneling each time the
  Bragg resonance is crossed, with a remaining fraction $1-r$ of the
  condensate being accelerated
  further. Agreement with theory is also good when, instead of changing the
  acceleration, the lattice
  depth is varied at fixed acceleration~(Fig.~\ref{fig4}(c)).

  Similar experiments were performed in the geometry
  leading to the larger lattice constant of $1.56\,\mathrm{\mu m}$. Because
  of the reduced Bloch
  velocity $v_{\rm B}$ in this geometry, the acceleration process was
  extremely sensitive
  to any initial velocity of the condensate, which in our TOP trap is
  intrinsically given by the
  micromotion~\cite{prlpaper} at the frequency of the bias field. For the
  trap parameters used in our
  experiments, the velocity amplitude of the micromotion could be of the same
  order of magnitude as
  $v_{\rm B}$ and the condensates could, therefore, have quasimomenta close
  to the edge of the
  Brillouin zone. In fact, when the standing wave was flashed on as in the
  Rabi-oscillation
  measurement described above, Bragg diffraction could be observed for zero
  detuning between the
  lattice beams, with the diffraction efficiency depending on the initial
  velocity of the condensate.
  Moreover, an initial velocity close to the band edge would have made it
  impossible to switch on the
  lattice adiabatically. In order to counteract these effects, we
  performed the acceleration
  experiments
  inside the magnetic trap, eliminating the velocity of the condensate
  relative to the lattice by
  phase-modulating one of the lattice beams at the same frequency and in
  phase with the rotating bias
  field of the TOP trap. In this way, in the rest frame of the lattice the
  micromotion was
  compensated. Nevertheless, a residual sloshing of the condensate with
  amplitudes $<3\,\mathrm{\mu m}$
  could not be ruled out, so that the uncertainty in the initial velocity of
  the condensate was still
  around $0.5\,\mathrm{mm\,s^{-1}}$, corresponding to $\approx
  0.3\,v_{\rm B}$ in this geometry. In
  Fig.~\ref{fig3} (c), the results of the acceleration of the condensate
  with a nominal lattice
  depth of $~1.38\,E_{\rm B}$ are shown together with the theoretical
  curves for
  $U_0=1.38\,E_{\rm B}$ and the (assumed) effective potential $U_{\rm
  eff}\approx0.65\,U_0\approx
  0.88\,E_{\rm B}$. The
  Landau-Zener tunneling probabilities measured in this lattice geometry
were compatible with
  the same
  effective potential assuming that the conversion factor
  between theoretical and actual lattice depth was the same as in the
  counter-propagating geometry
  in which $U_{\rm eff}\approx U_0$. In order to demonstrate unequivocally
  the reduction of the effective
  lattice potential by the interaction term, however, it would be necessary
  to vary $n_0$ appreciably
  holding all other parameters constant, which in our experimental setup was
  not possible without
  creating systematic errors due to variations in the equilibrium position of
  the trap (and hence the
  local laser intensity in the Gaussian profile) when the trap frequency was
  changed. As the
  interaction term is expected to distort the band structure of the
  condensate in the
  lattice~\cite{sorensen98}, it should affect all measurable quantities (Rabi
  frequency, amplitude of
  Bloch oscillations, and tunneling probablility~\cite{choi99}) in the same
  way, so that a
  differential measurement is necessary (as has been demonstrated in the pulsed
  Bragg-diffraction regime of Ref.~\cite{stenger99}). On the
  theoretical side, the finite extent of the condensate leading to
  the occupation of only a few lattice sites and the three-dimensional nature
   of the condensate evolution as well as the role of the interaction term
in the adiabaticity criterion for switching on the lattice
   will also have to be
  taken into account.

  In summary, we have investigated the coherent acceleration of
  Bose-Einstein condensates adiabatically loaded into a 1-D optical lattice
  as well as Bloch
  oscillations and Landau-Zener tunneling out of the lowest Bloch band. The
  results obtained are in
  good agreement with the available theories and extend the corresponding
  work on ultra-cold atoms in optical
  lattices into the domain of Bose-Einstein condensates.

  This work was
  supported by the MURST through the PRIN2000 Initiative, and by the
  EU through the Cold Quantum-Gases Network, contract
  HPRN-CT-2000-00125. O.M. gratefully acknowledges a
  Marie-Curie Fellowship from the EU within the IHP Programme.


\begin{references}
\bibitem{footnote} Similar expressions for the interaction
  parameter have been derived in~\cite{trombettoni01,chiofalo00}.
  The large lattice constant in the angle-geometry allowed us to
  realize a value of $C$ larger by a factor more than 10 with respect
  to ~\cite{anderson98} with a
  comparable  condensate density. On the
  other hand, the large lattice constant also means that only a
  few potential wells are occupied by the condensate.
\bibitem{footnote2} We have verified that in the process of the
acceleration and tunneling,
   the condensed fraction is not reduced. Our investigation did not,
however, test the evolution
  of the condensate phase, but on the basis of the Bragg scattering
experiments of~\cite{kozuma99}
  we assume that the interaction times of our
  experiment should not destroy the condensate phase.
\bibitem{footnote3}These expressions are only exact if the
  final velocity of the lattice is an integer multiple of $v_{\rm
  B}$, as in that case the mean velocity of the condensate is
  independent of the band structure.
\end{references}
  \end{document}